\documentclass{aastex6}
\fullcollaborationName{The Friends of AASTeX Collaboration}

\begin{document}


\title{Acceleration of Cooling of Ice Giants by Condensation in Early Atmospheres}


\author{Kenji Kurosaki\altaffilmark{1,2}}

\and

\author{Masahiro Ikoma\altaffilmark{1,3}}

\altaffiltext{1}{Department of Earth and Planetary Science, Graduate School of Science, The University of Tokyo, 7-3-1 Hongo, Bunkyo-ku, Tokyo 113-0033, Japan}
\altaffiltext{2}{
Department of Physics, Nagoya University, Furo-cho, Chikusa-ku, Nagoya-shi, Aichi 464-8602, Japan
}
\altaffiltext{3}{Research Center for the Early Universe, Graduate School of Science, The University of Tokyo, 7-3-1 Hongo, Bunkyo-ku, Tokyo 113-0033, Japan}
\email{KK: kurosaki.k@nagoya-u.jp; MI: ikoma@eps.s.u-tokyo.ac.jp}

\begin{abstract}
The present infrared brightness of a planet originates partly from the accretion energy that the planet gained during its formation and hence provides important constraints to the planet formation process. 
A planet cools down from a hot initial state to the present state by losing energy through radiative emission from its atmosphere. 
Thus, the atmospheric properties affect the planetary cooling rate.
Previous theories of giant planet cooling
assume that the atmospheric composition is unchanged throughout evolution. 
Planet formation theories, however, suggest that the atmospheres especially of ice giants are rich in heavy elements in the early stages. Those heavy elements include condensable species such as H$_2$O, NH$_3$, and CH$_4$, which are expected to have a great impact on atmospheric temperature and, thus, radiative emission through latent heat release.
In this study we investigate the effect of such condensation on the planetary emission flux and quantify the impact on the cooling timescale.
We then demonstrate that the latent heat of those species keeps the atmosphere hot and thus the emission flux high for billions of years, resulting in acceleration of the cooling of ice giants.
This sheds light on the long-standing problem that Uranus is much less bright than theoretically predicted and is different in brightness from Neptune in spite of similarity in mass and radius.
Also, we find that young ice giants with highly enriched atmospheres are much brighter in mid-infrared than those with unenriched atmospheres. This provides important implication for future direct imaging of extrasolar ice giants.

\end{abstract}


\section{Introduction} \label{sec:intro}

Infrared brightness of giant planets is often used for understanding how they formed and evolved to their present states and what has happened in their interior. 
During the formation, giant planets gain huge amounts of gravitational energy and then evolve by losing it into space via radiative emission. 
Thus, giant planets are initially hot, bright, and large in size, and subsequently become cool, dark, and small with age. 
By comparing the observed amount of infrared emission (hereafter the infrared luminosity) with theoretical luminosity that one obtains by integrating planetary thermal evolution over the age, one sometimes realizes that the theoretical model includes some wrong assumptions, which results in new findings. 

The solar-system gas giants Jupiter and Saturn have been relatively well investigated. 
The first theoretical model of Jupiter's thermal evolution \citep{Hubbard1977}, 
which assumed wholly convective and compositionally homogeneous interior, 
yielded successfully the infrared luminosity (or effective temperature) and radius that were consistent with their observed values. 
If applying the same model to Saturn, however, one obtains much lower luminosity than observed. 
This discrepancy between theory and observation led to the finding that helium is immiscible with hydrogen and settles down under high pressure in the Saturn's interior \citep[][and references therein]{Fortney2010}.

Also, recent observations have measured infrared luminosities of extrasolar gas giants \citep[e.g.,][]{Marley2007, Kuzuhara2013}.
As for such exoplanets detected by direct imaging, their masses are not measured but constrained by comparison between observed and theoretical luminosities, unlike solar-system planets. 
Through such studies regarding self-luminous gas giants, we have learned that it is crucial to understand thermal evolution of planets precisely for clarifying their origin and properties. 

In this study, our focus is on ice giants. 
The solar-system ice giants Uranus and Neptune are quite similar in mass and radius with each other, indicating that they have almost the same bulk composition. 
Nevertheless, Uranus is fainter in infrared by a factor of more than 10 than Neptune \citep{Pearl1990}. 
Also, the Uranus' faintness in itself is puzzling: 
Theoretical evolution models with the assumption of chemically homogeneous, wholly convective interior predict that Uranus is currently much brighter in infrared than observed \citep[e.g.,][]{Hubbard1980,Fortney2011,Nettelmann2016}. 
Those problems remain unsolved for more than 30 years (see section~\ref{sec:uranus} for the details).
Outside the Solar System, Neptune-class planets are quite common, according to recent exoplanet statistics \citep[e.g.,][]{Batalha2013}. 
Understanding the formation processes of those planets is, thus, a major issue in astronomy and planetary science. 
Regarding Neptune-class exoplanets, while infrared direct imaging has not been successful yet, 
future telescopes \citep[e.g., WFIRST-AFT,][]{Spergel2015arXiv}
are expected to detect infrared emission of cool ones like the ice giants in the Solar System.
To derive planetary properties from such observations, we need accurate knowledge of the evolution of ice giants.

Theoretical models of the thermal evolution of ice giants so far assume that the material distribution in the interior and atmosphere is unchanged from birth until the present. 
Recent theories of planet formation, however, suggest that the interior and atmosphere are more mixed in the formation stage. 
For example, small-size bodies such as pebbles and collisional fragments may contribute to the accretion of ice giant cores \citep{Inaba2003,Kobayashi2011,Lambrechts2014}. 
Such small bodies are evaporated in the envelopes and may accelerate formation of giant planets \citep{Hori2011,Venturini2015}
Also, Neptune-size planets are likely to experience giant collisions with protoplanets \citep{Slattery1992}, which would modify the internal compositional distribution greatly.
Since ice giants have large amounts of icy components such as H$_2$O, NH$_3$, and CH$_4$ in the interior, those icy components in the interior can be redistributed to the atmosphere by giant impacts.

This study is aimed to quantify the effect of icy-component enrichment in the atmosphere on the thermal evolution of ice giants. 
We assume that the ice giant possesses a significant amount of icy components in the atmosphere initially.
Because of their latent heat, condensation of those components is expected to affect the temperature gradient in the convective region of the atmosphere and raises the atmospheric temperature, 
as compared to that determined by dry adiabat.
Here we demonstrate that such a change in atmospheric temperature is effective in accelerating the cooling of ice giants significantly.

Condensation of the icy components may prevent convection in the hydrogen-dominated atmospheres like those of ice giants. 
This is because those condensable molecules are all heavier than hydrogen, so that condensation yields a mean molecular weight gradient in the atmosphere. 
This, however, remains a controversial issue. 
According to  \citet{Guillot1995-cond} and \citet{Leconte2016}, the mixing ratios of CH$_4$ in the atmospheres of Uranus and Neptune are higher than the threshold values beyond which convection is inhibited, but CH$_4$ clouds, which form via convective motion, are observed in the atmospheres \citep{Lindal1987, Orton2014a, Orton2014b}. 

The rest of this paper is organized as follows. 
In section~\ref{BS_U},
we describe our theoretical models and calculation methods of the atmosphere and interior structures and thermal evolution of ice giants. 
In section~\ref{result_U}, we show our calculation results, 
in particular, the effects of latent heat of icy components on the atmospheric structure and thermal evolution of the ice giant.
Then, in section~\ref{dis_U}, applying our new evolution models,
we discuss the implication for the faint Uranus paradox (\S~\ref{sec:uranus}) and the detectability of young exo-Neptunes (\S~\ref{sec:exoplanet}). 
Also, we discuss cavities of our evolution model including inhibition of convection caused by possible mean-molecular-weight gradients and resultant layered convection (\S~\ref{sec:cavities}).
Lastly, we summarize and conclude this paper in section~\ref{conc}.

\section{Model description} \label{BS_U}
In this study, we numerically integrate the thermal evolution of an ice giant, including the effect of condensation in its atmosphere. 
We assume that the planet consists of four layers in spherical symmetry and hydrostatic equilibrium, 
which include, from top to bottom, (1) an atmosphere composed of hydrogen, helium, water, ammonia, and methane, 
(2) an envelope composed of hydrogen, helium, and water, 
(3) a water-ice mantle and (4) a rocky core. 
At each interface, the pressure and temperature are continuous. 
We sometimes refer to the last three layers collectively as the interior, hereafter. 
Each layer in the interior is assumed to be fully convective and isentropic, and uniform in elemental abundance. 

\subsection{Interior structure}
The structure of the interior is determined by the differential equations,
\begin{eqnarray}
\frac{\partial P}{\partial M_r} &=& -\frac{GM_r}{4\pi r^4}, \label{IE1_U}\\
\frac{\partial r}{\partial M_r} &=& \frac{1}{4\pi r^2\rho},\label{IE2_U} \\
\frac{\partial T}{\partial M_r} &=& -\frac{GM_r}{4\pi r^4}\frac{T}{P}\nabla, \label{IE3_U}
\end{eqnarray}
and the equation of state,
\begin{equation}
\rho = \rho (P, T, X_i), \label{IE4_U}
\end{equation}
where $r$ is the planetocentric distance, 
$M_r$ is the mass contained in the sphere of radius $r$, 
$P$ is the pressure, $\rho$ is the density, $T$ is the temperature, 
$X_i$ symbolically denotes the composition,
and 
$G$ is the gravitational constant $(=6.67\times 10^{-8}$~dyn~cm$^2$~g$^{-2}$). 
The symbol $\nabla$ is the adiabatic temperature gradient with respect to pressure; namely,
\begin{equation}
\nabla = \nabla_{\mathrm{ad}} = \left(\frac{\partial \ln T}{\partial \ln P}\right)_S
\end{equation}

As for the equations of state, we use \citet{Saumon1995} for hydrogen and helium, SESAME 7150 for water \citep{SESAME}, 
and \citet{Valencia2007} for rock. For mixing of hydrogen, helium, and water in the envelope, we adopt the volume additive law \citep{Saumon1995},
\begin{equation}
\frac{1}{\rho} = (1-Z_0) \left( \frac{X}{\rho_{\mathrm{H}}} + \frac{Y}{\rho_{\mathrm{He}}} \right) + \frac{Z_0}{\rho_{\mathrm{H}_2\mathrm{O}}},
\end{equation}
where $X$, $Y$, and $Z_0$ are the mass fractions of hydrogen, helium, and water, respectively, and $\rho_\mathrm{H}$, $\rho_{\mathrm{He}}$, and $\rho_{\mathrm{H}_2\mathrm{O}}$ are the partial densities of hydrogen, helium, and water, respectively. We set $X=0.72$ and $Y=0.28$, which are the protosolar mass fractions without the heavy elements \citep{Lodders2009}. Since the behavior of ammonia and methane at high pressures and high temperatures is poorly known, we substitute the equation of state of water for those of ammonia and methane in accordance with previous researches \citep[e.g.,][]{Nettelmann2013-UN}. In this study, we deal with $Z_0$ as the sum of the fractions of water, ammonia, and methane.

\subsection{Atmospheric structure}
The atmosphere is assumed to be plane parallel and in the radiative-convective equilibrium. Thus, the net flux is constant through the atmosphere. The atmosphere contains H$_2$, He, H$_2$O, NH$_3$, and CH$_4$ gases and H$_2$O, NH$_3$, and CH$_4$ condensates (hereafter, H$_2$O, NH$_3$, and CH$_4$ are referred to collectively as the volatile). The distribution of each volatile component is determined by the phase equilibrium condition (i.e., saturation pressure). The interface between the atmosphere and interior (i.e., the bottom of the atmosphere) is assumed to be at the pressure level of 100~bars, which is denoted by $P_{\mathrm{btm}}$. 
The transition pressure $P_\mathrm{btm}$ must be high enough that condensation occurs in the atmosphere, not in the interior. On the other hand, since we assume that the mass of the atmosphere is negligibly small, the transition pressure must be low enough. We have chosen the value of 100 bar, because this value fulfills those conditions. The choice of transition pressure scarcely affects the results and conclusions given below, as long as those assumptions are valid.

The temperature-pressure relation in the radiative region (or stratosphere) is determined by the analytical formula derived by \citet{MatsuiAbe1986}: 
\begin{eqnarray}
\sigma T^4 & = & F_{\mathrm{top}} \frac{\tau + 1}{2} \nonumber \\
&&+ \frac{\sigma T_{\mathrm{eq}}^4}{2} \left[ 1+ \frac{\kappa_{\mathrm{th}}}{\kappa_{\mathrm{v}}} + \left( \frac{\kappa_{\mathrm{v}}}{\kappa_{\mathrm{th}}} - \frac{\kappa_{\mathrm{th}}}{\kappa_{\mathrm{v}}}  \right) \exp^{-\tau_{\mathrm{v}}} \right] \label{Eq_atm_str}
\end{eqnarray}
where $F_\mathrm{top}$ is the net flux, $T_{\mathrm{eq}}$ is the equilibrium temperature, $\kappa_{\mathrm{th}} $ and $\kappa_{\mathrm{v}} $ are the mean opacities for long- and short-wavelength radiation, respectively, $\tau$ and $\tau_{\mathrm{v}}$ are the optical depths for long- and short-wavelength radiation, respectively, and $\sigma$ is the Stefan- Boltzmann constant (= $5.67\times 10^{-5}~\mathrm{erg}~\mathrm{cm}^{-2}~\mathrm{K}^{-4}~\mathrm{s}^{-1}  )$.
The optical depth $\tau$ is defined as $d\tau = -\rho\kappa_{\mathrm{th}} dr$.

We assume that $\kappa_{\mathrm{th}}$ is the Rosseland mean opacity and $\kappa_{\mathrm{v}}=0.1\kappa_{\mathrm{th}}$. 
This assumption is verified by comparison between the above analytical formula and  $T$-$P$ profiles obtained from detailed radiative transfer calculations for hot Jupiter atmospheres \citep{Guillot2010}. 
We also neglect the short-wavelength radiation reflectivity and assume that the single scattering albedo is unity. 
By comparing our atmospheric model with that including the effects of short-wavelength stellar irradiation in detail \citep{Marley1999}, we have confirmed that short-wavelength radiation has little influence on the temperature-pressure structure of $\tau > 1$, which is of special interest in this study. 
The opacities of H$_2$ and He are due to collision induced absorption and Rayleigh scattering, for which
we have used the data table from \citet{Freedman2008}: 
The data are available for density $\rho =2.5\times 10^{-12}$ to $10$~g~cm$^{-3}$ and temperature $T=10^2$ to $10^4$~K. 
The opacities of H$_2$O, NH$_3$, and CH$_4$ gases are calculated from line profiles derived from the HITRAN 2012 database \citep{Rothman2013}. 
The calculation method of line profiles is based on \citet{Rothman1998}. 
We adopt the Voigt profile as the line shape and use the analytical formula that \citet{Kuntz1997} derived by implementing Humlices's algorithm for approximation. \citet{Kuntz1997} showed that the implementation of Humlicek's algorithm yields errors of at most $2\times10^{-6}$, which stems from the cutoff for the Voigt function. 
The monochromatic opacity $\kappa(\nu)$ is calculated as 
\begin{equation}
\kappa(\nu) = \frac{k_{\eta\eta'} (\nu, T, P)}{M},
\end{equation}
where $M$ is the mass of the molecule and $k_{\eta\eta'} $ is the monochromatic absorption coefficient for the transition between lower $\eta$ and upper $\eta'$ states. 
We obtain the Rosseland mean opacities of H$_2$ $(\kappa_{\mathrm{H_2}})$, 
He $(\kappa_{\mathrm{He}})$, H$_2$O $(\kappa_{\mathrm{H}_2\mathrm{O}})$, 
NH$_3$ $(\kappa_{\mathrm{NH}_3})$, and CH$_4$ $(\kappa_{\mathrm{CH}_4})$ molecules from line intensities. 
Then, we calculate the total opacity
$\kappa_\mathrm{th}$ as
\begin{eqnarray}
\kappa_\mathrm{th} &=& x_\mathrm{H_2} \kappa_{\rm H_2} + x_\mathrm{He}\kappa_{\rm He}  
\nonumber \\
& &
+ x_\mathrm{H_2O}\kappa_{\rm H_2O} + x_\mathrm{NH_3}\kappa_{\rm NH_3} + x_\mathrm{CH_4}\kappa_{\rm CH_4} \label{kap_atm_nu}
\end{eqnarray}
where $x_\mathrm{H_2}$, $x_\mathrm{He}$, $x_\mathrm{H_2O}$, $x_\mathrm{NH_3}$, and $x_\mathrm{CH_4}$ are the mole fractions of H$_2$, He, H$_2$O, NH$_3$, and CH$_4$, respectively.

The temperature gradient in the convective region (or troposphere) is determined by the pseudo-moist adiabatic one. For $N$ kinds of species including $j$ kinds of non-condensable species, the pseudo-moist adiabatic temperature gradient is given by \citep{Ingersoll1969,Atreya1986, Abe1988}
\begin{equation}
\frac{d\ln T}{d\ln P} = \nabla_\mathrm{dry}
\frac{1+\sum_{i=j+1}^{N} \frac{x_i}{1-x_{i}} \frac{d\ln~p_{i}^{\ast}}{d\ln~T} }
{1+\sum_{i=j+1}^{N} \frac{R_g}{C_p} \frac{x_i}{1-x_{i}} \frac{d\ln~p_{i}^{\ast}}{d\ln~T}}, \label{Conv_atm_multi}
\end{equation}
where $\nabla_\mathrm{dry}$ is the adiabatic temperature gradient without condensation (i.e., dry adiabatc), 
$C_p = \sum_{i=1}^{N} x_i C_{p,i} $ is the mean heat capacity, $x_i$ and $p_{i}^{\ast}$ are the mole fraction and vapor pressure of condensable species $ i~(i = j+1,\cdots, N)$. 
The vapor pressure of H$_2$O and those of NH$_3$ and CH$_4$ are taken from \citet{Nakajima1992} and \citet{Sanchez2004}, respectively. 
We assume that the heat capacities of H$_2$O, NH$_3$, and CH$_4$ at constant pressure are $4~R_g$, where $R_g$ is the gas constant, as an ideal gas approximation. The atmospheric temperature of interest in this study is less than 500~K for H$_2$O and less than 300~K for NH$_3$ and CH$_4$. 
In such a temperature range, the assumption for the heat capacity is in good agreement with experiments. 
We ignore the non-ideal effect of the condensates, because it is negligible at the height of  $\tau \sim 1$, where the radiation from the top of the atmosphere is determined.

We integrate the radiation transfer equation by using the Eddington approximation. The upward and downward radiation flux densities $F_{\mathrm{IR}}^{+}$ and $F_{\mathrm{IR}}^{-}$ can be written as
\begin{eqnarray}
F_{\mathrm{IR}}^{+}(\tau) &=& \pi B(\tau) - \int_{\tau_b}^{\tau} \frac{d}{d\tau'} ( \pi B(\tau') )
\exp\left( -\frac{3}{2}(\tau' - \tau) \right) d\tau', \label{Fu} \\
F_{\mathrm{IR}}^{-}(\tau) &=& \pi B(\tau) - \int_{0}^{\tau} \frac{d}{d\tau'} ( \pi B(\tau') )
\exp\left( -\frac{3}{2}(\tau-\tau') \right) d\tau'  \nonumber \\
&& - \pi B(0) \exp\left( -\frac{3}{2}\tau \right), \label{Fd} 
\end{eqnarray}
the net radiative flux as
\begin{equation}
F_{\mathrm{rad}} = F_{\mathrm{IR}}^{+} - F_{\mathrm{IR}}^{-} - F_{\mathrm{irr}}, \label{Frad}
\end{equation}
and  the net flux as
\begin{equation}
F_{\mathrm{net}} = F_{\mathrm{rad}}  + F_{c}  \label{Fnet},
\end{equation}
where $B (\tau)$ is the blackbody radiation intensity, $F_c$ is the convective flux and $F_{\mathrm{irr}}$ is the direct solar flux. Note that $F_{\mathrm{top}}=F_{\mathrm{IR}}^{+} (\tau=0)$. We assume the net flux is constant through the atmosphere and the convective flux is equal to zero in the stratosphere.

\subsection{Thermal evolution}
For simulating the thermal evolution, we integrate the energy conservation equation, 
\begin{equation}
\frac{\partial L_r}{\partial M_r} = -T\frac{\mathrm{d}S}{\mathrm{d}t} \label{therm-1}
\end{equation}
where $S$ is the specific entropy, $L_r$ is the total energy flux passing through a sphere of radius $r$, and $t$ is time. Integrating the equation from the center to the top, we obtain 
the intrinsic luminosity $L_\mathrm{int}$ as
\begin{eqnarray}
L_{\mathrm{int}} & = & - \left[ \frac{\mathrm{d}S_{\mathrm{env}}}{\mathrm{d}t} \int_{M_c+M_m}^{M_p} TdM_r \right. \nonumber \\
&& \left. + \frac{\mathrm{d}S_{\mathrm{m}}}{\mathrm{d}t} \int_{M_c}^{M_c+M_m} TdM_r +\frac{\mathrm{d}S_{\mathrm{c}}}{\mathrm{d}t} \int_{0}^{M_c} TdM_r \right], \label{Eq_thermal_U}
\end{eqnarray}
where $S_{\mathrm{env}}$, $S_m$, and $S_c$ are the specific entropies of the envelope, the mantle, and the core, respectively, and $M_c$ and $M_m$ are the masses of the core and mantle, respectively. $L_{\mathrm{int}}$ is also written as $L_{\mathrm{int}}=4\pi R_p^2 F_{\mathrm{top}}$ where $R_p$ is the planetary radius and $F_{\mathrm{top}}$ is the outgoing flux from the top of the atmosphere.

\subsection{Numerical procedure}
Numerically we integrate Eq.~(\ref{Eq_thermal_U}) by the following procedure. 
First, we determine the atmospheric structure and the intrinsic luminosity $L_{\mathrm{int}}$, which correspond to the outer boundary conditions for the interior structure. 
To do so, assuming trial values of $F_{\mathrm{top}}$ and the pressure $P_{\mathrm{ad}}$ at the boundary between the stratosphere and the troposphere, which we call the tropopause pressure, 
we calculate the $T$-$P$ profile in the stratosphere from Eq.~(\ref{Eq_atm_str}) and that in the troposphere from Eq.~(\ref{Conv_atm_multi}). 
From those profiles, we calculate the upward and downward fluxes from Eqs. (\ref{Fu})--(\ref{Fnet}). The tropopause pressure $P_{\mathrm{ad}}$ is determined by the divergence of the net flux for radiation. 
That is, in the troposphere, the net flux satisfies
\begin{equation}
 \frac{dF_{\mathrm{rad}}}{d\tau}  \sim   \frac{F_{\mathrm{rad}}^{i+1} - F_{\mathrm{rad}}^{i}}{\tau_{i+1}-\tau_i} < 0
\end{equation}
while in the stratosphere, 
\begin{equation}
 \frac{dF_{\mathrm{rad}}}{d\tau} \sim  \frac{F_{\mathrm{rad}}^{i+1} - F_{\mathrm{rad}}^{i}}{\tau_{i+1}-\tau_i} \ge 0, 
\end{equation}
where $i$ is the grid number. 
We assume that the top of the atmosphere is $P_{\mathrm{top}}$ = $1 \times 10^{-5}$~bars. 
If the tropospheric temperature derived by Eq.~(\ref{Eq_atm_str}) is smaller (higher) than that derived by Eq.~(\ref{Conv_atm_multi}), the trial value of $P_{\mathrm{ad}}$ is higher (smaller) than the actual value. 
Then, we set the value of $P_{\mathrm{ad}}$ for the next step smaller (higher). 
Also, we replace the value of $F_{\mathrm{top}}$ with the value calculated by Eq.~(\ref{Fu}). 
We repeat this procedure until we obtain the self-consistent values of $P_{\mathrm{ad}}$ and $F_{\mathrm{top}}$ that satisfies the relative error of temperature at the tropopause smaller than 1~\%. 
At this point, we determine the temperature at $P = P_{\mathrm{btm}}$ (denoted by $T_{\mathrm{btm}}$) from which we calculate the specific entropy of the envelope. 

We determine the masses of the rocky core ($M_c$), the water mantle ($M_m$), and the envelope ($M_e$) from the total mass fractions of heavy elements in the interior $(Z_{\mathrm{tot}})$ and in the envelope $(Z_0)$, and the water mass fraction in the interior $(Z_w)$, which are respectively defined as
\begin{eqnarray}
Z_{\mathrm{tot}} &=& \frac{M_{e,w}+M_m+M_c}{M_p}, 
\\
Z_0 &=& \frac{M_{e,w}}{M_e}, 
\\
Z_w &=& \frac{M_{e,w}+M_m}{M_p}, 
\end{eqnarray}
where $M_{e,w}$ is the total mass of water in the envelope.
That is, we calculate the masses fractions of the core, mantle, and envelope as 
\begin{eqnarray}
\frac{M_c}{M_p} &=& (1-Z_w)\cdot Z_{\mathrm{tot}}, \\
\frac{M_m}{M_p} &=& Z_w\cdot Z_{\mathrm{tot}}- \frac{Z_0 (1-Z_{\mathrm{tot}})}{1-Z_0 }, \\
\frac{M_e}{M_p} &=&1-Z_{\mathrm{tot}}+\frac{Z_0 (1-Z_{\mathrm{tot}})}{1-Z_0}, 
\end{eqnarray}
respectively.

Finally, equation~(\ref{Eq_thermal_U}) is numerically integrated in the following way. 
For each time step, we first simulate two interior models for two different values of $T_{\mathrm{btm}}$. 
We integrate Eqs.~(\ref{IE1_U})-(\ref{IE4_U}) inward from the atmospheric bottom, $M_r = M_p$ (neglecting the mass of the atmosphere), to the planetary center, $M_r = 0$, using the fourth-order Runge-Kutta method. 
We then look for the solution that fulfills the inner boundary condition (i.e., $r = 0$ at $M_r = 0$) in an iterative fashion, by changing the planetary radius ($R_p$). 
Note that determining $P_{\mathrm{ad}}$ and $F_{\mathrm{top}}$ requires the gravity in the atmosphere (or the planetary radius $R_p$), which is obtained after the interior structure is determined. Thus, we have to find the solution in which the interior and atmospheric structures are consistent with each other also in an iterative fashion. Then we calculate the time interval $\Delta t$ from the second-order difference equation for Eq. (\ref{Eq_thermal_U}), which is written as
\begin{eqnarray}
	\Delta t = - \frac{C}{L_\mathrm{int} (t+\Delta t) + L_\mathrm{int} (t)}
	\label{NM01_U}
\end{eqnarray}
where 
\begin{eqnarray} 
	C & = & \left. [ S_{\mathrm{env}} (t+\Delta t) - S_e (t)  ][ \Theta_{\mathrm{env}} (t+\Delta t) + \Theta_\mathrm{env} (t) ]\right. \nonumber\\
&&\left. +[ S_{\mathrm{m}} (t+\Delta t) - S_m (t)  ][ \Theta_{\mathrm{m}} (t+\Delta t) + \Theta_m (t) ] \right. \nonumber \\
&&\left. +[ S_c (t+\Delta t) - S_c (t)  ][ \Theta_c (t+\Delta t) + \Theta_c (t) ]\right. ,
\end{eqnarray}
\normalsize and  
\begin{eqnarray}
	\Theta_{\mathrm{env}} (t) & \equiv & \int^{M_p}_{M_m+M_c} T (t) dM_r, \\
	\Theta_{\mathrm{m}} (t) & \equiv & \int^{M_m+M_c}_{M_c} T (t) dM_r, \\
	\Theta_c (t) & \equiv & \int^{M_c}_{0} T (t) dM_r,
	\label{NM02_U}
\end{eqnarray}
and 
$S_\mathrm{env}$, $S_m$, and $S_c$ are the entropies of the envelope, the mantle, and the core, respectively.
We have confirmed that our numerical code for the atmospheric structure reproduces well the $P_{\mathrm{ad}}$ and $F_{\mathrm{top}}$ values presented by \citet{Nakajima1992}. Also, we have confirmed that our numerical code for the interior structure reproduces well the mass and radius relationship for super-Earths presented by \citet{Valencia2010}.

The free parameters in this model are the total mole fraction of the volatiles (i.e., H$_2$O, NH$_3$, and CH$_4$) in the atmosphere $x_{\mathrm{btm}} (t = 0)$, 
the initial mole ratios of NH$_3$/H$_2$O and CH$_4$/(H$_{2}+$He) at the atmospheric bottom, 
the abundance of helium with respect to hydrogen ($Y$), 
the pressure at the atmospheric bottom $P_{\mathrm{btm}}$, 
the total heavy elements mass fraction of the planet $Z_{\mathrm{tot}}$, 
and the total mass fraction of heavy elements in the envelope $Z_0$.
In this study, we assume $Z_{w} = 0.95$. 
The conversion relation between $x_\mathrm{btm}$ and $Z_0$ is
\begin{equation}
x_{\mathrm{btm}} = \frac{Z_0}{\frac{\mu_W}{\mu_H}(1-Z_0)+Z_0}, \label{xbtm_Z0}
\end{equation}
where 
$\mu_\mathrm{W}$ and $\mu_\mathrm{H}$
are the molecular weights of H$_2$O (= 18.0) and 
of the mixture of H$_2$ and He with solar abundances (= 2.3), respectively.
We assume that the abundance of water in the envelope is constant through the evolution.
We also assume the initial mole ratios of NH$_3$/H$_2$O $=0.135$, which is equal to the N/O value of the solar abundance \citep{Lodders2009},
and the mole fraction of CH$_4$/(H$_{2}+$He)$=0.023$, which is the present CH$_4$ mixing ratio of Uranus's atmosphere \citep{Lindal1987, Marley1999}.
Table~\ref{paramTbl} summarizes the parameters and their fiducial values.

Finally, since the atmosphere cold enough for the volatiles to condense is of interest in this study,  we adopt the Uranus' value (58.2~K) for $T_\mathrm{eq}$.
The incident solar radiation flux ($L_{\mathrm{eq}}$) is calculated by $L_{\mathrm{eq}}=4\pi R_p^2\sigma T_{\mathrm{eq}}^4$. 
Also, we assume 
the planetary mass is Uranus' mass (= $8.68\times 10^{28}$~g).

\begin{table}[tbp]
\begin{center}

\begin{tabular}{c|c|c|c}
$M_p$ [g] & $T_{\mathrm{eq}}$ [K] & $Z_W$ & $Z_{\mathrm{tot}}$ \\
\hline
$8.68\times10^{28}$ & 58.2 & 0.95 & 0.90 \\
\hline
\hline
 $P_{\mathrm{btm}}$ [bar] & $x_{\mathrm{btm}} $ [mol~\%] &  NH$_3$/H$_2$O  & CH$_4$/(H$_{2}+$He) \\
\hline
 100 & 2.3-50 & 0.135 & 0.023
\end{tabular}

\caption{\label{paramTbl} Parameters used in the fiducial result.}
\end{center}
\end{table}

\section{Results} \label{result_U}
\subsection{Atmospheric structure}
\begin{figure}[htbp]
\begin{center}
\includegraphics[width=7cm]{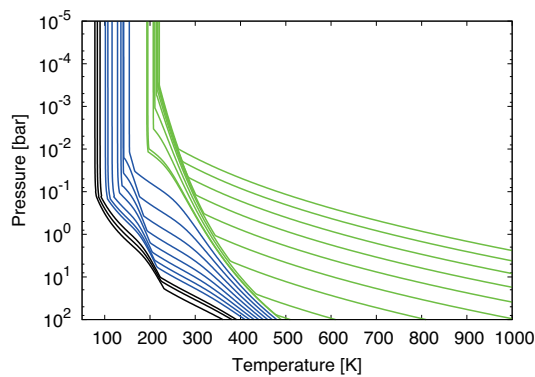} 
\caption{\label{Atm_str} 
Temperature-pressure profiles in the atmosphere for different values of the temperature at 100~bar, $T_\mathrm{btm}$.
The atmosphere consists of H$_2$, He, H$_2$O, NH$_3$, and CH$_4$, the mole fractions of which are 45.6~\%, 4.42~\%, 43.0~\%,  5.80~\%, and 1.18~\%, respectively. 
Condensation of H$_2$O, NH$_3$, and CH$_4$ are taken into account.
The green, blue, and black lines represent the cases where the main condensate is H$_2$O, NH$_3$, and CH$_4$, respectively.
The equilibrium temperature is assumed to be 58.2~K.
}
\end{center}
\end{figure}

First we investigate the effect of condensation on the atmospheric structure.
Figure~\ref{Atm_str} shows $T$-$P$ profiles in the atmosphere for various values of $T_{\mathrm{btm}}$.
We set the mole fractions of H$_2+$He, H$_2$O, NH$_3$, and CH$_4$ are 50.0~\%, 43.0~\%, 5.80~\%, and 1.18~\%, respectively.
The atmospheric structure is composed of three parts, a dry convective region, a moist convective region, and a radiative region (i.e., a stratosphere) from the bottom to the top.
The temperature gradient in the dry convective region is determined by the heat capacity 
of the dominant constituents,
whereas it is controlled by the latent heat of the condensable constituents in the moist convective region.
The green lines in Fig.~\ref{Atm_str} represent the case where the main condensed constituent
is H$_2$O. 
As $T_\mathrm{btm}$ decreases, the $T$-$P$ curve shifts from right to left. 
In Fig.~\ref{Atm_str}, it turns out that the shift does not occur uniformly. 
In the lower atmosphere, temperature at a given pressure changes with $T_\mathrm{btm}$, whereas it is almost unchanged in the middle and upper atmosphere. This is because condensation of H$_2$O occurs in the middle atmosphere, so that the $T$-$P$ curve is fixed to the saturation $T$-$P$ relation (i.e., vapor pressure).  Once almost all of water is depleted by precipitation, the $T$-$P$ profile moves leftwards as a whole. Then, the next condensate (i.e., NH$_3$ and CH$_4$ in the blue-line and black-line cases, respectively) determines the $T$-$P$ profile in the middle atmosphere.

\begin{figure}[htbp]
\begin{center}
\includegraphics[width=7cm]{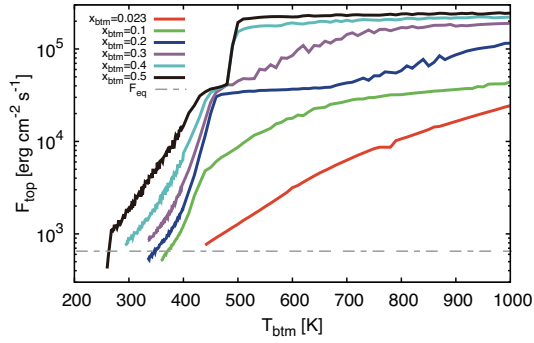} 
\caption{\label{T_Ftop} 
Calculated relationship between the outgoing radiative flux at the top of the atmosphere
$F_{\mathrm{top}}$ and the temperature at 100~bar in the atmosphere $T_\mathrm{btm}$ for 
six different values of the total mole fraction of the volatiles (i.e., H$_2$O, NH$_3$, and CH$_4$)
$x_{\mathrm{btm}}=$ 0.023 (red), 0.1 (green), 0.2 (blue),  0.3 (purple), 0.4 (cyan), and 0.5 (black). 
The horizontal thin black line represents
the equilibrium flux $\sigma T_\mathrm{eq}^4$.
Here we have assumed the mole fraction of CH$_4$/H$_2$$=0.023$ and NH$_3$/H$_2$O$=0.135$.
and the equilibrium temperature is 58.2~K. 
}
\end{center}
\end{figure}

Figure~\ref{T_Ftop} shows the outgoing flux from the top of the atmosphere $F_{\mathrm{top}}$ as a function of $T_{\mathrm{btm}}$ for different values of the total mole fraction of the volatiles $x_{\mathrm{btm}}$.
To see the condensation effect on the outgoing flux, we first focus on the result for $x_{\mathrm{btm}}=0.5$ (black line).
For $T_\mathrm{btm} \gtrsim$ 500~K, $F_\mathrm{top}$ is nearly constant (= $3.3 \times 10^{5}~\mathrm{erg}~\mathrm{s}^{-1}\mathrm{cm}^{-2}$). 
This value is similar with the Komabayashi-Ingersoll limit of H$_2$O dominated atmospheres \citep{Nakajima1992}. 
That is because the atmospheric temperature structure is determined by the moist convection of H$_2$O, as described above.
Also, the surface of $\tau \simeq 1$, which we call hereafter the atmospheric photosphere, is in the moist convective region.
That is why $F_{\mathrm{top}}$ is nearly constant in the temperature range where the atmospheric structure is dominated by the moist convection of H$_2$O.
Hereafter we call such a constant $F_{\mathrm{top}}$ the radiation limit.

The radiation limit also appears when condensation of NH$_3$ or CH$_4$ occurs. 
The flux $F_{\mathrm{top}}$ changes sharply at $T_\mathrm{btm}$ $\simeq$~500~K.
Once $T_\mathrm{btm}$ reaches about 450~K,
$F_{\mathrm{top}}$ settles down to the radiation limit due to NH$_3$ condensation ($\sim 3 \times 10^{4}~\mathrm{erg}~\mathrm{s}^{-1}\mathrm{cm}^{-2}$). 
$F_\mathrm{top}$ decreases only slightly in the range of $T_\mathrm{btm} \simeq$ 420-450~K because of NH$_3$ condensation, and then decreases rapidly for $T_\mathrm{btm}$ $\lesssim$ 420~K after NH$_3$ is depleted. 
In the case of small $x_\mathrm{btm}$, 
the domain of slow change around $T_\mathrm{btm} \simeq$ 400~K disappears because of the low abundance of NH$_3$. 

Finally, for $T_\mathrm{btm} \lesssim$ 250~K, the atmospheric structure is affected by CH$_4$ condensation.
However,
since CH$_4$ is not abundant enough for its latent heat to dominate the structure,  
$F_{\mathrm{top}}$ is not constant but deceases, as $T_\mathrm{btm}$ decreases.

As $x_\mathrm{btm}$ decreases, $F_\mathrm{top}$ decreases for a given $T_\mathrm{btm}$. This is because the contribution of latent heat to heat capacity becomes small with decreasing $x_\mathrm{btm}$, so that the photospheric temperature decreases. Indeed, in the case of dry convection ($x_\mathrm{btm}$ = 0.023; red line), $F_\mathrm{top}$ is considerably small relative to, for example, that for $x_\mathrm{btm}$ = 0.5 (black line). 
Also, because of low abundances of the condensable molecules, the domains of constant $F_\mathrm{top}$ disappear in cases of low $x_\mathrm{btm}$.

\subsection{Thermal evolution}
\label{result_U2}
\begin{figure}[htbp]
\begin{center}
\includegraphics[width=7cm]{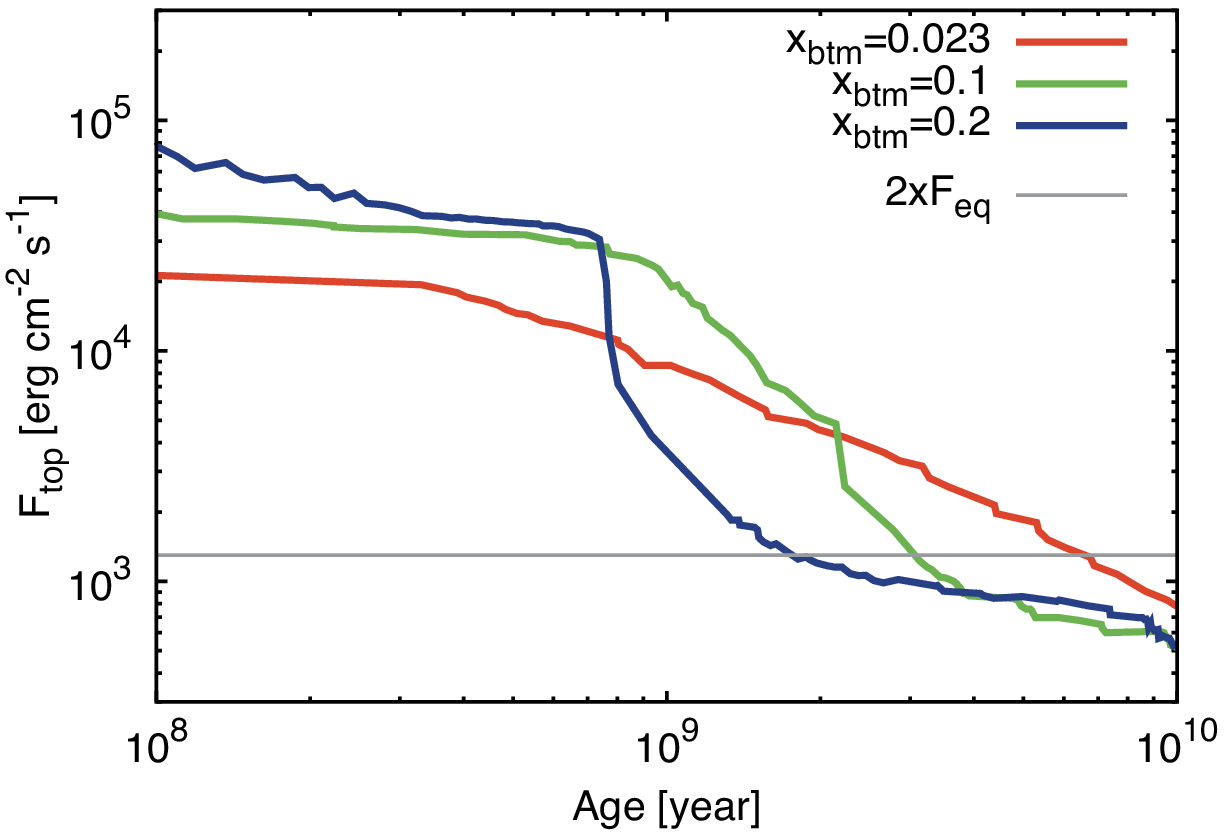} 
\includegraphics[width=7cm]{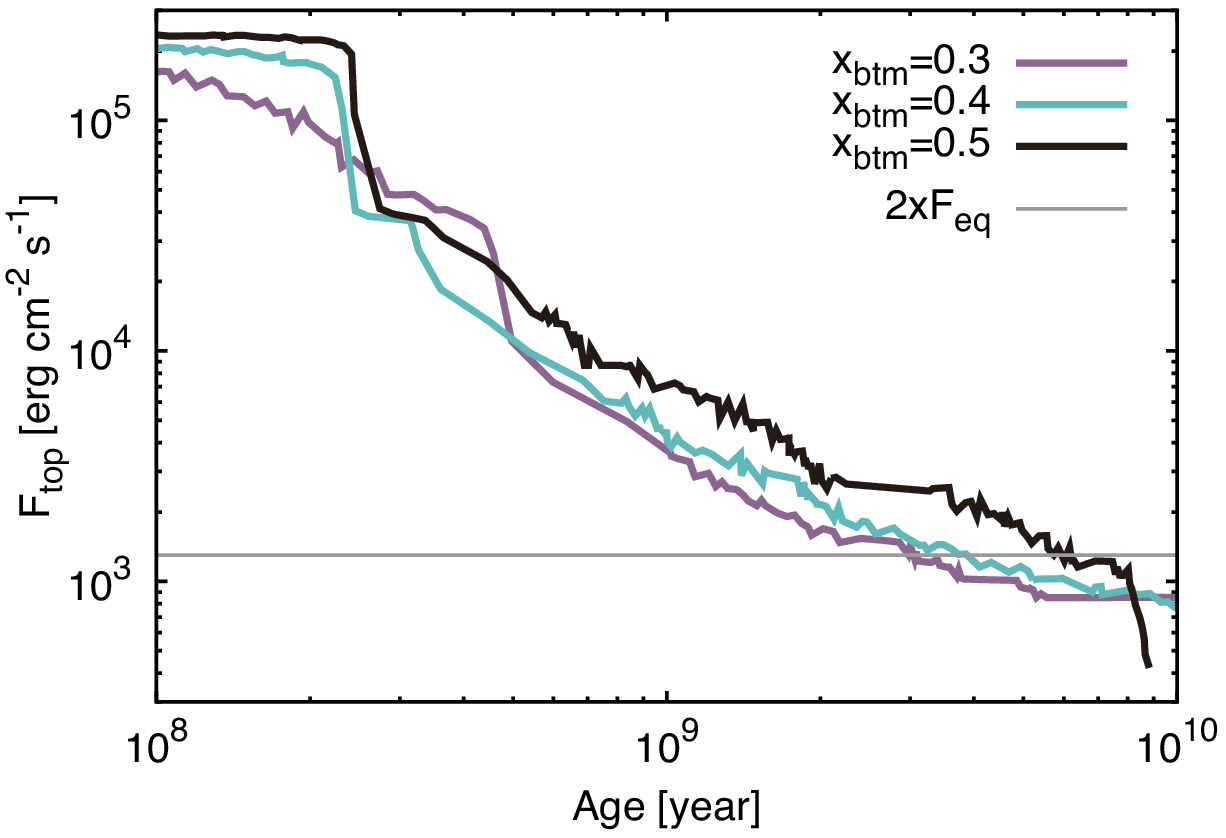} 
\caption{\label{Leff_Z0_sol} 
Evolution of the outgoing flux at the top of the atmosphere $F_{\mathrm{top}}$ for different values of the initial  total mole fraction of the volatiles (H$_2$O+NH$_3$+CH$_4$) in the atmosphere. 
Panels (a) and (b) show the results for 
$x_{\mathrm{btm}}$ = 0.023 (red), 0.1 (green), and 0.2 (blue) 
and those for $x_{\mathrm{btm}}$ = 0.3 (purple), 0.4 (cyan), and 0.5 (black), respectively. 
Here we have assumed that the equilibrium temperature $T_\mathrm{eq}$ is 58.2~K.
The grey line represents two times the incident stellar radiative flux, $F_{\mathrm{eq}}$ (i.e., $2 \sigma T_\mathrm{eq}^4$). 
The values of the other parameters used are listed in Table~\ref{paramTbl}.
}
\end{center}
\end{figure}

\begin{figure}[htbp]
\begin{center}
\includegraphics[width=7cm]{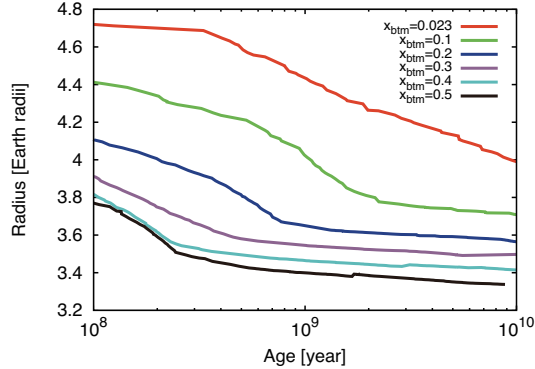}             
\caption{\label{Rp_Z0_sol}
Evolution of the planetary radius. 
Lines are colored according to the initial total mole fraction of the volatile (H$_2$O+NH$_3$+CH$_4$) in the atmosphere,  
$x_{\mathrm{btm}}=$ 0.023 (red), 0.1 (green), 0.2 (blue), 0.3 (purple), 0.4 (cyan), and 0.5 (black). 
Parameter values used for these calculations are the same as those used in Fig~\ref{Leff_Z0_sol}.
}
\end{center}
\end{figure}

Here we investigate the thermal evolution of ice giants with volatile-rich atmospheres.
Figure~\ref{Leff_Z0_sol} shows the evolution of $F_{\mathrm{top}}$ for different values of $x_\mathrm{btm}$.
The evolution of the planetary radius $R_p$ is also shown in Fig.~\ref{Rp_Z0_sol}.
As seen in Fig.~\ref{Leff_Z0_sol}, the planet with an atmosphere richer in the volatiles is more luminous in early stages. 
This is because the stratospheric temperature is kept high because of the release of latent heat in the troposphere (Fig.~\ref{Atm_str} and \ref{T_Ftop}).
If the atmosphere is more enriched initially, $F_\mathrm{top}$ is larger and the cooling occurs more rapidly,
so that the effective temperature reaches the equilibrium value earlier.
For example, $F_\mathrm{top}$ reaches $2\times F_{\mathrm{eq}}$ within $4 \times10^{9}$~years if $x_{\mathrm{btm}}$ = 0.1 to 0.4{, whereas it takes 7$\times 10^{9}$ years if $x_\mathrm{btm}$ = 0.023}.
This is because,
as shown in Fig.~\ref{T_Ftop}, larger $x_\mathrm{btm}$ causes larger $F_\mathrm{top}$, leading to rapid cooling of the planet.
However, $F_\mathrm{top}$ for $x_\mathrm{btm}=0.5$  is large compared to that for $x_\mathrm{btm} \leq$ 0.4 at $t >$ 4~Gyr.
That is because the condensation of CH$_4$ dominates the planetary flux in the case of $x_\mathrm{btm} =$~0.5.

Figure~\ref{Rp_Z0_sol} shows the evolution of the planetary radius for the same parameter values as those used in Fig~\ref{Leff_Z0_sol}. 
The larger $x_\mathrm{btm}$ is, the smaller the radius is at a given age for two reasons: First, because large $x_\mathrm{btm}$ corresponds to heavy-element-rich interior (i.e., large $Z_0$; see Eq.~[\ref{xbtm_Z0}]), the planet is small in size. 
Second, as seen in Fig.~\ref{Leff_Z0_sol}, large $x_\mathrm{btm}$ results in rapid cooling of the planet, which leads to rapid contraction. 
Because of those two effects, the planet with a volatile-rich atmosphere is considerably small in size relative to that with a volatile-poor atmosphere.

\section{Discussion}\label{dis_U}
\subsection{Faint Uranus Paradox \label{sec:uranus}} 

The faint Uranus paradox is a long-standing unsolved problem. 
As mentioned in Introduction, Uranus is much darker in infrared than theoretically predicted. 
More exactly, if integrating the thermal evolution of Uranus backwards from the present state, based on conventional models that assume the material distribution in the interior and atmosphere has been the same as the present \citep{Hubbard1980,Fortney2011},
one obtains the initial effective temperature of as low as 65-70~K. 
This temperature corresponds to the planet accretion timescale of the order of ten billion years, which means that Uranus' formation timescale is much longer than the age of the Solar System. 

A few ideas have been proposed that heat transport is prevented in the interior for some reason, resulting in low emission.
One possible mechanism is inefficient convection due to compositional gradient \citep{Stevenson1982,Podolak2012}, which causes the so-called double diffusive convection.
The double diffusive convection, however, tends to disappear in about one billion years and, instead, overturning vigorous convection occurs in the interior \citep{Kurokawa2015}.
Another possible mechanism is delay in cooling due to the presence of thermal boundary layers in the interior.
For example, the transition region between the H/He-dominated envelope and the ice-dominated mantle is stably stratified and behaves as a thermal boundary layer.
\citet{Nettelmann2016} demonstrated that the thermal boundary layer could work well to reproduce Uranus's low luminosity. The problem remains, however, of why the same thing does not occur in Neptune.

\begin{figure}
\begin{center}
\includegraphics[width=7cm]{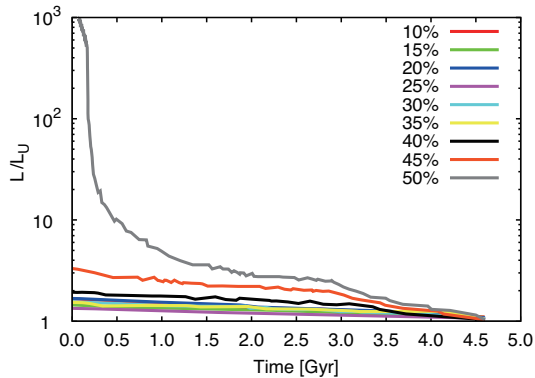}
\caption{\label{Uranus_ev}
Evolution of the planetary luminosity $L$ in the unit of the present Uranus' luminosity $L_\mathrm{U}$ (= $5.6 \times 10^{22}~\mathrm{erg}~\mathrm{s}^{-1})$. 
Lines are colored according to the total mole fraction of the volatiles (i.e., H$_2$O, NH$_3$, and CH$_4$) in the atmosphere from 10-mol~\% to 50-mol~\%. 
We assume the total heavy element in the planet $Z_\mathrm{tot}=0.9$ 
and the mass fraction of water $Z_\mathrm{W}=0.95$.
The heavy element in the planetary envelope $Z_0$ is determined by $x_\mathrm{btm}$ using Eq.~(\ref{xbtm_Z0}).
$Z_\mathrm{tot}, Z_\mathrm{W}$, and $Z_0$ are constant through the evolution.
The planetary mass is $8.68\times 10^{28}$~g.
}
\end{center}
\end{figure}

Here, applying our new evolution model, we propose a scenario that can explain the Uranus' infrared darkness and the difference from Neptune. 
In Fig.~\ref{Uranus_ev}, we show the thermal evolution models that we have obtained by integrating the  evolution backwards from the present (the present luminosity $L_\mathrm{U}$ being $5.6\times 10^{22}~\mathrm{erg}~\mathrm{s}^{-1}$).  
The values of parameters other than $x_\mathrm{btm}$ are given in Table~\ref{paramTbl}.
Our calculation demonstrates that the planet with an atmosphere of $x_\mathrm{btm} =$ 50~mol~\% is quite luminous at $t = 0$.
Namely, if Uranus' atmosphere had been sufficiently polluted with condensable constituents initially,
planetary cooling would have occurred rapidly enough to result in the present luminosity of Uranus within the age of the solar system. 

The questions are what caused the atmospheric pollution and why such a process occurred only for Uranus.
One possible process is giant impacts.
Namely, a giant impact dredged up condensable constituents from the icy deep interior, distributing them around the proto-Uranus to form a circumplanetary disk.
Then the atmosphere became highly polluted due to re-accretion of the condensable constituents from the circumplanetary disk.
This scenario would be also consistent with the highly inclined obliquity of Uranus (= 97.77$^{\circ}$), which can be explained by a giant impact event \citep{Slattery1992, Morbidelli2012}.
On the other hand, Neptune may have not experienced such a giant impact, because of its low obliquity (= 28.32$^{\circ}$).
If that is the case, Neptune had a less polluted, dry atmosphere initially, which also explains the present Neptune's luminosity \citep{Fortney2011,Nettelmann2013-UN}.
Thus, giant impacts likely played a key role in creating the big difference in brightness between Uranus and Neptune.

However, the planetary radius that we have calculated here is smaller than the present radius of Uranus (= 4.0~$R_\oplus$).
In the case of $x_\mathrm{btm}$ = 50-mol~\% ($Z_0$ = 0.887),
the calculated present planetary radius is 3.4~$R_{\oplus}$.
In order to produce
both the planetary radius and intrinsic luminosity that are the same as those of the present Uranus,
we have to assume the atmosphere is more polluted than the envelope ($x_\mathrm{btm}$ = 0.5 and $Z_0$ = 0.3, for example), which is hydrodynamically unstable, in principle. 
In the above scenario, icy materials are supplied to the geometrically thin atmosphere from the circumplatetary disk. The hydrodynamic stability of such a heavy, thin layer and
the efficiency of ice material transport between the atmosphere and interior are to be examined by hydrodynamic simulations.
Thus, for the faint Uranus paradox being resolved within the scenario of accelerated cooling due to atmospheric pollution, the origin and stability of such distribution of the volatiles remain crucial issues to be explained by future studies.

Note that the above hypothesis for the origin of the two different ice giants does not necessarily mean that there is a dichotomy between those two classes of ice giants in extrasolar systems.
Planetary luminosity is a function of time, planetary mass, planetary radius and composition. 
Even for planets with the same age, mass, and radius, the atmospheric composition will differ from planet to planet. This is partly because the outcome of a giant impact differs greatly, depending on the collisional angle and velocity. Thus, a continuum of luminosity, not a dichotomy, will be seen in extrasolar systems.

\subsection{Self-luminous extrasolar ice giants} \label{sec:exoplanet}
As mentioned in Introduction, future observation is expected to detect the thermal emission of extrasolar ice giants directly. 
In this study we have found that ice giants, if their atmospheres being polluted with the volatiles, are much more luminous in early stages than predicted without atmospheric pollution.
Thus, it would be interesting to quantify how large difference the effect of atmospheric pollution yields in the planet-to-star contrast.

\begin{figure}
\begin{center}
\includegraphics[width=7cm]{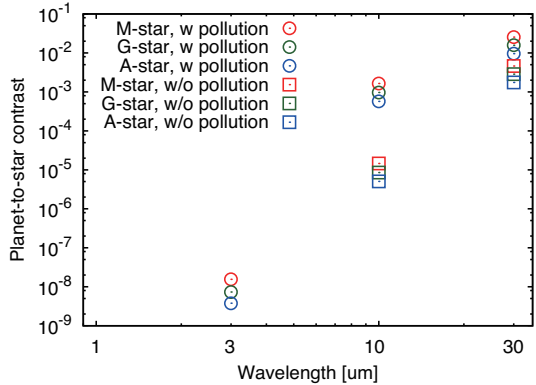}
\caption{\label{obs_fig} 
Predicted brightness contrast between the ice giant and the host star of age 0.1~Gyr (see eq.~[\ref{eq:contrast}]). The assumed mass and equilibrium temperature of the ice giant are the same as those of Uranus (see Table~1).
The open circles and squares represent the results with and without atmospheric pollution ($x_\mathrm{btm}$ = 0.023 and 0.5), respectively. 
The symbols are colored according to stellar type, M-type (0.3~$M_\odot$), G-type (1.0~$M_\odot$), and A-type (1.8~$M_\odot$) main-sequence stars. 
Note that small values of the contrast, $C$, at 1 and 3 $\mu$m are not presented: For the non-polluted cases, $C \sim$ $10^{-15}$ and $10^{-42}$ at 1 and 3~$\mu$m, respectively; $C \sim$
$10^{-23}$ at 1~$\mu$m for the polluted case.
}
\end{center}
\end{figure}

Here we estimate the contrast $C$ as
\begin{equation}
C = \frac{\int_{\Delta\lambda} B_{\lambda}(T_{\mathrm{eff}}) d\lambda} 
              {\int_{\Delta\lambda} B_{\lambda} (T_{\ast}) d\lambda},
\label{eq:contrast}
\end{equation}
where $B_\lambda$ is the Planck function for wavelength $\lambda$, $\Delta \lambda$ is the bandwidth, $T_\mathrm{eff}$ and $T_\ast$ are the effective temperatures of the planet and the host star, respectively.
Figure~\ref{obs_fig} shows the estimated value of $C$ for the ice giant at age 0.1~Gyr in the cases without pollution ($x_\mathrm{btm}$ = 0.023; open circles) and with pollution ($x_\mathrm{btm}$ = 0.5; open squares). The values of the other parameters are given in Table~1. We have consider different types of host star, which include M-type ($0.3~M_{\odot}$; red symbols), G-type ($1.0~M_{\odot}$; green symbols), and A-type ($1.8~M_{\odot}$; blue symbols).
The evolution of the host star is calculated with the SSE package \citep{Hurley2000}. 
The wavelength $\lambda$ is assumed to be 1, 3, 10, and 30 $\mu$m with bandwidth $\Delta\lambda=0.2~\mu$m.

Since planets are cooler than host stars, the contrast becomes higher, in general, as wavelength becomes longer. 
Thus, longer wavelength is preferable for direct detection. 
Also, since the polluted atmosphere of the ice giant is hotter ($\sim$ 250~K) than the unpolluted atmosphere ($\sim$ 140~K) in early stages, the contrast for the former (open circles) is higher than that for the latter (open squares). Furthermore, the contrast difference between the polluted and unpolluted atmospheres decreases, as wavelength increases, as shown in Fig.~\ref{obs_fig}. 
Namely, for distinguishing between the two types of atmosphere, intermediate wavelengths are preferable. 
Indeed, in Fig.~\ref{obs_fig}, it is found that the difference is about two orders of magnitude at $\lambda$ = 10~$\mu$m; $C \sim 10^{-5}$ for the unpolluted case, whereas $C \sim 10^{-3}$ for the polluted case. 
It is interesting to note that the detection level for high contrast imaging with MIRI on JWST is $\sim 10^{-4}$ for 1~arcsec \citep{Beichman2010}. 
While evaluating the actual detectability via direct imaging is beyond the scope of this study, 
this simple analysis done here would be sufficient to indicate it is quite important to do detailed investigation. 

\subsection{Cavities} \label{sec:cavities}
In this study we have made some simplifications. 
First, while we have assumed that the structure is completely adiabatic (or iso-entropic) in the convective regions of the atmosphere, it is not exactly adiabatic but super-adiabatic, in reality. 
In particular, if the compositional gradient is large enough to
prevent convection from transporting entropy efficiently from the deep interior, 
the temperature gradient is super-adiabatic, 
which was pointed out by \citet{Guillot1995-cond} in the case of the giant planets.
Inefficient convections that delays planetary cooling include layered convection, in which mode small convective cells are separated by diffusive interfaces \citep{Radko2003, Rosenblum2011, Mirouh2012,Wood2013}.
The transport efficiency of the layered convection depends on the layer thickness. 
The impact of the layered convection on the thermal evolution was recently examined in the case of gas giants by \citet{Vazan2015}, \citet{Kurokawa2015} and \citet{Nettelmann2015}.
They showed that the layered convection was hydrodynamically unstable against large-scale overturning convection,  
resulting in smoothing out the compositional inhomogeneity in the planetary interior. 
Also, as mentioned in Introduction, although theories predict convection should be inhibited in Uranus' and Netpune's atmospheres \citep{Guillot1995-cond,Leconte2016}, 
CH$_4$ clouds, which are formed by convection, are observed there. 
Our assumption of pseudo-moist convection in the atmosphere is supported by the recent hydrodynamic study by 
\cite{Li2015} who showed that the atmospheric temperature gradient was determined approximately by the pseudo-moist adiabat. 
Their calculation is, however, only for low mole fractions of ice constituents. 
Similar investigations is to be done for validating our assumption.

Second, we have ignored the effect of clouds on the thermal structure of the atmosphere.
Since clouds absorb infrared radiation, addition of clouds makes the atmosphere optically thicker, which causes the pressure at the photosphere becomes lower. 
The temperature at the photosphere, which is in the moist-convective region, is determined by the vapor pressure. Thus, the presence of clouds lowers the photospheric temperature and thus
the emitted flux from the photosphere.
Consequently, the cooling time is prolonged due to the effect of clouds.
Clouds also affect the planetary radius, because the presence of clouds increases the thickness of the atmosphere and delays the planet cooling. 
The former effect is tiny, because the height of the cloud level corresponds to only about 0.1 \%  of the planetary radius. 
Also we have confirmed that the latter effect is not great enough to reproduce the present Uranus's radius. 
If the present Uranus's luminosity is reproduced, the atmosphere contains 50 mol~\% of ice constituents (see Fig.~\ref{Uranus_ev}). 
However, as shown in Fig.~\ref{Rp_Z0_sol}, the planetary radius varies at most 10~\% between $10^8$ and $10^{10}$ years. Thus, the radius is at most $3.8 R_{\oplus}$, which is smaller than Uranus.
The detailed treatment of clouds is beyond the scope of this study.
However, since the effect would have a non-negligible impact on the thermal evolution of ice giants, 
this is an important issue to be addressed in future studies.

\section{Conclusions} \label{conc}
Here we have quantified the effect of latent heat for condensation of H$_2$O, NH$_3$, and CH$_4$ in the atmosphere on the thermal evolution of ice giants for the first time.
We have found out that the latent heat keeps the atmosphere warm and the planetary thermal emission high.
Integrating the thermal evolution of ice giants with that effect, we have demonstrated that ice giants with atmospheres highly enriched in those condensable species are significantly bright in early evolution stages and thereby evolve on short timescales, which means that the aged ice giants are less bright than predicted by previous theories.
We have also found that the evolution timescale is sensitive to the amount of condensable constituents in the atmosphere.
Our finding of such rapid thermal evolution sheds light on the long-standing issue that Uranus is much less bright than theoretically predicted and is different in brightness from Neptune in spite of similarity in mass and radius.
Also, our finding that young ice giants with enriched atmospheres are quite luminous is important for detecting extrasolar ice giants by direct imaging.

\acknowledgments

We thank Y. Abe and K. Hamano for fruitful discussion about the atmospheric structure. We also thank Y. Sekine for constructive comments on the manuscript. 
We also thank the anonymous referee for his/her careful reading and constructive comments that helped us improve this paper greatly.
This work is supported partly by JSPS Core-to-Core Program "International Network of Planetary Sciences" and 
Grant-in-Aid for JSPS Research Fellow (No. 26-11515; K.K.), Scientific Research (A) (No. 16H02160; K.K.), for Scientific Research on Innovative Areas (No. 23103005; M.I.) and Scientific Research (C) (No. 25400224; M.I.).



\listofchanges

\end{document}